\documentclass[3p, twocolumn]{elsarticle}
\usepackage{lineno,hyperref}
\usepackage{graphicx}
\usepackage{xcolor}
\usepackage{amsmath}
\usepackage{comment}

\modulolinenumbers[5]

\journal{Physics Letters B}

\begin{document}

\begin{frontmatter}

\title{Bayesian Neural Networks for Fast SUSY Predictions}
\author[1]{B. S. Kronheim}
\ead{brkronheim@davidson.edu}
\author[1]{M. P. Kuchera}
\ead{mikuchera@davidson.edu}
\author[2]{H. B. Prosper}
\author[1]{A. Karbo}
\address[1]{Department of Physics, Davidson College, Davidson, NC 28035, USA}
\address[2]{Department of Physics, Florida State University, Tallahassee, FL 32306, USA}


\begin{abstract}
One of the goals of current particle physics research is to obtain evidence for new physics, that is, physics beyond the Standard Model (BSM), at accelerators such as the Large Hadron Collider (LHC) at CERN. The searches for new physics are often guided by BSM theories that depend on many unknown parameters, which, in some cases, makes testing their predictions difficult. In this paper, machine learning is used to model the mapping from the parameter space of the phenomenological Minimal Supersymmetric Standard Model (pMSSM), a BSM theory with 19 free parameters, to some of its predictions. 
Bayesian neural networks are used to predict cross sections for arbitrary pMSSM parameter points, the mass of the associated lightest neutral Higgs boson, and the theoretical viability of the parameter points. All three quantities are modeled with average percent errors of 3.34\% or less and in a time significantly shorter than is possible with the supersymmetry codes from which the results are derived. These results are a further demonstration of the potential for machine learning to model accurately the mapping from the high dimensional spaces of BSM theories to their predictions.
\end{abstract}


\end{frontmatter}


\section{Introduction}

The discovery of the Higgs boson at the LHC in 2012\,\cite{Aad:2012tfa,201230} marked the end of the search for the Standard Model (SM) particles. With the completion of the SM, physicists' focus over the next decade or so is understanding the physics of electroweak symmetry breaking\,\cite{EWSB} by following two broad strategies: comparing precision measurements of Higgs boson properties with SM predictions\,\cite{DAWSON20191} and conducting direct searches for physics beyond the SM (BSM). This marks a methodological change in particle physics, moving from a well-posed search for particles predicted by a well-tested theory to searching for any evidence of new physics guided, in part, by the predictions of BSM theories. A popular group of candidate theories for BSM physics are the supersymmetric (SUSY) theories. These theories provide potential solutions to the hierarchy problem, permit gauge coupling unification at high energies\,\cite{hep-ph/9309277}, and provide a promising candidate for a dark matter 
particle\,\cite{JUNGMAN1996195}. 

The simplest formulation of supersymmetry consistent with the SM is the Minimal Supersymmetric Standard Model (MSSM). The MSSM uses the same gauge group as the SM and assumes minimal particle content and R-parity conservation. Despite being a minimal model, the MSSM has 105 free parameters\,\cite{Djouadi:376049} beyond those of the SM, making a thorough exploration of the model challenging. The typical approach is to select a set of parameter points within an accessible subset of the parameter space and compute observables, such as cross sections, for each point. While meaningful results have been obtained from approaches like this\,\cite{1308.0297}, the expensive computations limit our ability to investigate the theoretical parameter spaces thoroughly and limit our ability to use standard likelihood methods\,\cite{likelihood} to make inferences about the parameter spaces. Through the methods outlined in this paper, we show that an accurate, fast mapping of BSM theory parameters to predictions can be constructed based on recently available tools that implement sampling via Hamiltonian Monte Carlo (HMC).

A simplification of the full MSSM, which is also a prototypical example of a BSM theory 
is the phenomenological MSSM (pMSSM) \cite{Ambrogi:2017lov,Fawcett:2016xoh,Aad:2015baa,Khachatryan:2016nvf,Sekmen:2011cz,Berger:2008cq}. The pMSSM has no new sources of CP-violation, no flavor changing neutral currents, and includes first and second generation universality. These assumptions, which are consistent with experimental facts, reduce the 105 free parameters to just 19 \cite{Berger:2008cq}. The large reduction in the number of free parameters is useful in that it renders calculations with this model feasible, while being large enough to make the pMSSM a good proxy of the MSSM. The parameter space is also complex enough to highlight the advantages of the use of machine learning for the rapid calculation of observables. While this is the theory studied in this paper, the pMSSM is merely an interesting example of a high-dimensional theory 
that illustrates the proposed technique for fast predictions.

Machine learning has been successfully applied to several problems in high energy physics\,\cite{Carleo:2019ptp} and SUSY in particular. In \cite{PhysRevD.98.052004}, neural networks were shown to be capable of determining restrictions on BSM parameter spaces given experimental data and in \cite{Caron2017} random forests were used to classify pMSSM parameter points as excluded or not excluded by ATLAS and CMS searches. In \cite{bechtle2017}, neural networks are used to calculate the profile likelihood ratios for a variation of the pMSSM, while in \cite{Caron:2019xkx}, an alternative to random sampling, called active learning, is used to explore the pMSSM parameter space. Additionally, Bayesian neural networks, the type of machine learning used in this paper, along with boosted decision trees, were used in \cite{Abazov:2009ii,Aaltonen:2009jj,topQuarks} to aid in the detection of single top quarks at the Tevatron as well as in neutrino background and signal discrimination \cite{Xu_2008}.
For recent reviews of the use of machine learning in the physical sciences see, for example, \cite{Carleo:2019ptp,Bhat:2010zz} and the recently released machine learning inference toolkit \texttt{MadMiner} \cite{Brehmer:2019xox}.

In this paper, we use Bayesian neural networks\,\cite{neal_1996} to model the mapping of the parameter space of the pMSSM to its predictions. The program \texttt{SOFTSUSY} \cite{ALLANACH2002305} is used to calculate particle spectra and decay chains, while the program \texttt{Prospino2} is used to calculate cross sections for neutralino chargino  pair production\,\cite{Production} at the LHC at 14 TeV\,\cite{Schmidt:2016jra}. These 
programs encode algorithms that give accurate predictions, but only point by point in
the parameter space. However, given predictions computed at a large number of parameter
points, we show how Bayesian neural networks (BNNs) can be used to create prediction functions that map parameters to predictions. This is demonstrated for three different prediction functions:
\begin{enumerate}
        \item A map from parameters to a classification of whether a given pMSSM parameter point is physically or numerically viable as determined by \texttt{SOFTSUSY}.
        \item A map from parameters to the predicted cross section for neutralino chargino production.
        \item A map from parameters to the predicted light neutral Higgs boson mass.
    \end{enumerate}
With these functions, it is possible to assess quickly whether a pMSSM parameter point is valid, whether it yields a neutral Higgs boson mass consistent with the observed value, and predict the cross section for neutralino chargino production. 


\section{Mathematical Details}
 Our goal is to predict the physical or numerical viability of pMSSM parameter points, predict the neutralino chargino production cross section, and predict the mass of the lightest neutral Higgs boson, and to do so as  accurately and rapidly as possible. The pMSSM parameters are listed in Table\,1. 
\begin{table}[ht]
\begin{center}
\begin{tabular}{ | p{1.5cm} | p{2.8cm} | p{2.3cm} | } 
\hline
Parameter & Description & Range \\ 
\hline
$M_1$ & bino mass & $|M_1| \leq$ 4 TeV \\ 
\hline
$M_2$ & wino mass & $|M_2| \leq$ 4 TeV \\ 
\hline
$M_3$ & gluino mass & $M_3 \le$ 4 TeV \\ 
\hline
$\mu$ & higgsino mass & $|\mu|\leq$ 4 TeV \\
\hline
$M_A$ & pseudoscalar Higgs boson mass & $M_A \le$ 4 TeV \\ 
\hline
$\tan\beta$ & ratio of vacuum expectation values of Higgs doublets  & 1 $\leq\tan\beta \leq$ 60\\ 
\hline
$A_t, A_b, A_{\tau}$ & third generation trilinear coupling & $A \leq$ 7 TeV \\ 
\hline
$m_{\bar{q}}$, $m_{\bar{u}_R}$, $m_{\bar{d}_R},$ $m_{\bar{l}}$, $m_{\bar{e}_R}$ & first/second generation sfermion mass parameters & $m \leq$ 4 TeV \\ 
\hline
$m_{\bar{Q}}$, $m_{\bar{t}_R}$, $m_{\bar{b}_R},$ $m_{\bar{L}}$, $m_{\bar{\tau}_R}$ & third generation sfermion mass parameters & $m \leq$ 4 TeV \\ 
\hline
\end{tabular}
\end{center}
\label{tab:bounds}
\caption{The 19 parameters of the pMSSM and the subset of the pMSSM parameter space considered in this paper.}
\end{table}
In this paper, we model these functions as BNNs\,\cite{neal_1996}.

\subsection{Bayesian neural networks}
In the Bayesian approach neural networks \cite{neal_1996}, the goal is to infer a probability density $p(\theta \, | \, D)$ over the parameter space of the network given training data $D$. For the pMSSM, the training data $D = \{(t_k, x_k)\}$ 
comprise targets $t_k$ associated with the pMSSM parameters points $x_k$. 
A Bayesian neural network is a functional 
\begin{align}
    p(x, D) & = \int F(x, \theta) \, p(\theta \, | \, D) \, d\theta
\label{eq:BNN}
\end{align}
of the posterior density
 \begin{align}
 p(\theta \, | \, D) & = \frac{p(D \, |\, \theta)\, \pi(\theta)}{p(D)}, 
 \label{eq:posterior}
 \end{align}
where $p(D \, | \, \theta)$ is the likelihood of the data, $\pi(\theta)$ a prior density, and $F(x, \theta)$ a function whose average over the space is desired. For example, setting $F(x, \theta) = \delta(y - f(x, \theta))$, where $f(x, \theta)$ is a neural network, yields the predictive density,
\begin{align}
    p(y \, | \, x, D) & = \int \delta(y - f(x, \theta)) \, p(\theta \, | \, D) \, d\theta .
    \label{eq:predictive}
\end{align}
In practice, the posterior
density is represented by an ensemble of neural networks whose parameters are sampled from
the posterior density using the Hamiltonian Monte Carlo method~\cite{neal_1996,Betancourt2017ACI}. Since the method approximates Eq.\,(\ref{eq:predictive}), it automatically furnishes an estimate of the uncertainty in the predictions $y$ from some measure of the width of the predictive density.

{\color{red}


}

\subsection{Likelihood and Prior}
\label{subsec:likelihood}
A crucial step in constructing a Bayesian neural network is modeling
the joint probability $p(t, x, \theta)$, which as noted above, is usually factorized into a
likelihood function $p(D | \theta)$ and a prior $\pi(\theta)$. The likelihood is the product  $\prod_j p(t_j|x_j, \theta) \, p(x_j | \theta)$ over all sampled pMSSM parameter points. However, if we make the reasonable assumption that the pMSSM parameters $x$ are independent of the network parameters $\theta$, that is, $p(x_j|\theta) = p(x_j)$, the likelihood function to be modeled is $p(t | x, \theta)$. 
\paragraph{Likelihood} 
For the classifier, we take the likelihood function $p(t | x, \theta)$ to be a Bernoulli density with targets 
$t = 1$ and 0 for the viable (\texttt{valid}) and non-viable (\texttt{invalid}) pMSSM parameter points, respectively. For the regression models, the targets are either the cross sections computed using \texttt{Prospino2} or the Higgs boson masses computed using \texttt{SOFTSUSY}. For regression, the likelihood function $p(t | x, \theta)$ is chosen to be a normal density $\mathcal{N}(t, f(x, \theta), \sigma)$ with mean $f(x, \theta)$ and an unspecified variance $\sigma^2$.

If one neglects theoretical uncertainties, 
codes such as \texttt{Prospino2} and \texttt{SOFTSUSY} provide deterministic predictions $t = P(x)$. Consequently, the interpretation of the probabilistic mapping from $x$ to $t$ is somewhat subtle.  Since the predictions are noiseless, strictly speaking the likelihood describing $t$ and $x$ is the $\delta$-function $\delta(t - P(x))$, which is approximated using $p(t | x, \theta)$ $=$ $\mathcal{N}(t, f(x, \theta),  \sigma)$.  

\paragraph{Prior}
Choosing a high-dimensional prior for likelihood functions is an extremely difficult problem. It is particularly challenging
for functions as complex as neural networks in which the parameters $\theta$ have no obvious meaning. For such cases, well motivated methods have been proposed to construct so-called objective priors (see, for example, \cite{Berger_2015}). However, these methods are computationally prohibitive for high-dimensional spaces and are not guaranteed to yield satisfactory results. 
Therefore, in practice, the prior is chosen for computational simplicity and its ability to yield satisfactory results. Furthermore, by using a hierarchical prior whose parameters are constrained by a hyper-prior, increased flexibility is introduced that makes it possible to tune the prior by varying the hyper-parameters in order to improve the quality of the results. \color{black} The overall prior $\pi(\theta)$ is a product of the priors for all network parameters and the associated hyper-priors. 

An obvious choice for a prior is a product of zero mean normal densities, one for each neural network parameter. This choice is equivalent to imposing $L_2$ regularization in standard neural network training (see, for example, \cite{DBLP:journals/corr/Schmidhuber14}). However, Neal \,\cite{neal_1996}  showed (see, also, \cite{lee2017deep}) that due to the central limit theorem Bayesian neural networks with normal priors tend to converge to a Gaussian process prior in the limit of an infinite number of hidden nodes. Since this is not necessarily the desired behavior, in this work we have chosen to use Cauchy priors, which alters the behavior of large Bayesian neural neural networks. For additional details about the prior used see~\cite{TensorBNN}.

There is another prior whose effect we need to consider. As noted in the next section, $x \sim p(x)$, where $p(x)$ is a flat prior in the bounded region of the pMSSM parameter space shown in Table\,1. 
Two possible concerns come to mind. The first is that the experimentally accessible region of the pMSSM may not be fully covered by the region listed in Table\,1. The second is that our results  may be sensitive to the choice of the sampling density $p(x)$. 


As noted above, the likelihood $p(D | \theta)$ depends on our model $p(x | \theta)$ for the pMSSM parameter space sampling density, $p(x)$. However, the posterior density over the network parameter space is given by
 $p(\theta | D) = \prod_j p(t_j | x_j, \theta) \, p(x_j | \theta) \, \pi(\theta) / p(D)$, which,
 given the assumption $p(x| \theta) = p(x)$ and noting that $p(D) \propto \prod_j  p(x_j)$, shows that the \emph{explicit} dependence on $p(x)$ drops out. Of  course, the implicit dependence on $p(x)$ remains through the specific training sample $\{ x_j \}$. As a consequence, the accuracy of results will depend on the sampling density of the training data. While a uniform sampling of the parameter space is the simplest to implement, it would clearly be better to place more points where they are needed most. 
 One way to do so would be to weight each pMSSM point $x_j$ by a likelihood function that incorporates LHC data.   
 
 In addition to the dependence of results on the training data, results also depend on the prior $\pi(\theta)$. In a more thorough study the effect of this prior on the inferences would be assessed by weighting every neural network in the ensemble by the ratio of the new prior to the one with which the ensemble was generated. The ability to do this in a practical way was not available at the time of writing, but is now available in the Bayesian neural network package used in this work \cite{TensorBNN}. 
 \color{black}

\section{Data Sets}
\label{sec:datasets}
In this section, we describe the data sets used to construct the prediction functions, that is, the functions mapping pMSSM parameter points to predictions.
\color{black}
We use three independent data sets, labeled \textbf{VPAR}, \textbf{OHIGGS}, \textbf{XSEC}, and follow standard practice by dividing each into three sets: training, validation, and test sets in the percentages, 80\%, 10\%, and 10\%, respectively. The first set was used to
train the BNN models, the second was used to assess the models' performance during training, and the third (the test set) was used to evaluate the performance of the trained models.

As noted above, our choice of $p(x)$ may miss subsets of the experimentally viable pMSSM points. But, in order to determine the boundary at which the experimental sensitivity drops below a given threshold, careful studies of the predicted signals and associated Standard Model backgrounds, using analyses optimized for different integrated luminosities, would be needed.  Such studies would be extremely interesting, but are beyond the scope of this paper. Instead, we restrict our attention to the region of the pMSSM parameter space
that has been used in other studies\,
\cite{Khachatryan:2016nvf,Cahill-Rowley:2014twa}. 
\color{black}
\paragraph{VPAR} This data set consists of 500,000 pMSSM points randomly sampled from the subspace given in Table\,1. Each pMSSM parameter is sampled independently from a uniform distribution over its range. For each point, \texttt{SOFTSUSY} is used to compute sparticle masses and decays. Of the points sampled, 60.61$\%$ were labeled as invalid and 39.39$\%$ as valid. 

\paragraph{OHIGGS} This data set consists of 567,597 points with sparticle masses computed using \texttt{SOFTSUSY}. The points were sampled in the same way as for \textit{VPAR}, but only those points flagged as valid were kept and that yielded the lightest neutral Higgs boson mass between 110 and 130 GeV, as this contained the majority of the points and had a much smaller range than the full dataset of  591,337 points.

\paragraph{XSEC} This data set consists of 202,264 pMSSM points with decays computed using \texttt{SOFTSUSY} and cross sections computed at next to leading order (NLO) accuracy using \texttt{Prospino2}. The \texttt{SOFTSUSY} calculations were reused from the generation of the \textit{OHIGGS} data set.  


\subsection{Data preparation}
The data sets were prepared for training using several different normalization schemes. For the input data, here the 19 pMSSM parameters, we scaled and shifted each parameter to have zero mean and unit variance. The targets of the \textbf{VPAR} data set were
left unchanged since the values are 0 and 1. The targets of the \textbf{XSEC} data set were log normalized, that is,  the natural logarithm of the cross sections 
was computed and the values shifted and scaled to have zero mean and unit variance. This was done because the distribution of the log of the cross section was roughly normal. The \textbf{OHIGGS} data set was also normalized to have zero mean and unit variance by subtracting the mean and dividing by the standard deviation.
\section{Training}
\subsection{Architecture and Training}
The Bayesian approach can be applied to any machine learning model. However, because of the computational burden of Markov chain Monte Carlo methods, even when aspects of the calculations can be parallelized,  the models used tend to be smaller than the ones trained (that is, fitted) using optimization methods such as stochastic gradient descent. 
A detailed technical description of 
the BNN implementation we have used, as well as details of the Hamiltonian Monte Carlo (HMC) sampling method, is given in\,\cite{TensorBNN} and the package developed for this work is available at \cite{TensorBNNcode}.
But, for completeness, we briefly describe the main features of the  models we used and their training.

Each model is a fully connected feed-forward neural network with 19 inputs, one for each pMSSM parameter, and 5 linear layers, each with 50 nodes. The output node is a sigmoid function for the classifier and is linear for the regression functions. The activation functions are variations on a ReLU\, \cite{TensorBNN}.
In order to provide a good starting point for the Hamiltonian Monte Carlo sampling, each model
was 
trained by minimizing the mean squared error for regression and the mean binary cross entropy for classification
with a batch size of 32. The
training was run for three cycles of 30 epochs each using AMSGRAD\,\cite{reddi2019convergence}. An epoch is a pass over the full training data while a cycle, in this context, is training with a given learning rate. For regression the learning rates were 0.01 for the first cycle, 0.001 for the second, and 0.0001 for the third, while for classification the learning rates were a factor 10 smaller. For each cycle, the network with the smallest validation error served as the starting point for the next cycle and the best network of the last cycle was used as the starting point for the HMC sampling. 

Network sampling was run 
until there was evidence of convergence, which we took to mean no statistically significant changes in the predictive distributions.
For example, for the cross section regression function the sampling was run for 13,500 epochs after a burn-in of 100 epochs. Here an epoch is a sequence of deterministic steps through the neural network parameter space governed by a simplectic approximation to Hamilton's equations. (For details, see\,\cite{TensorBNN}.) In order to produce an approximately \emph{iid}  sample of networks for inference, the original was down-sampled by 100, yielding a sample of size 135 for the cross section regression function, and by a factor 10 for the other two functions. For completeness, Table\,\ref{tab:BNNparams} lists the values of the hyper-parameters we used in the HMC sampling, but we defer to\,\cite{TensorBNN} for an explanation of the quantities listed.
\begin{table}[ht]

\begin{center}
\begin{tabular}{ | p{2.7cm} | p{0.95cm} | p{1.0cm} | p{1.45cm} |} 
\hline
Parameter & \textbf{VPAR} & \textbf{XSEC} & \textbf{OHIGGS}\\ 
\hline
leapfrog start & 2000 & 1000 & 1000 \\ 
\hline
leapfrog min & 2000 & 100 & 100 \\ 
\hline
leapfrog max & 10000 & 2000 & 2000 \\ 
\hline
step size start & 5e-5 & 1e-5 & 5e-5 \\ 
\hline
step size min & 2.5e-5 & 5e-6 & 2.5e-5\\ 
\hline
step size max & 2e-4 & 2e-5 & 1e-4\\ 
\hline
leapfrog grid step & 10 & 1 & 1 \\ 
\hline
hyper step size & 5e-6 & 1e-5 & 1e-5\\ 
\hline
output $\sigma$ & - & 0.1 & 0.5\\ 
\hline
epochs & 1750 & 13500 & 2850\\ 
\hline
\end{tabular}

\end{center}
\caption{\texttt{TensorBNN} network training parameters for each of the prediction functions. Explanations of the parameters may be found in\,\cite{TensorBNN}.}
\label{tab:BNNparams}
\end{table}

The training was done using an Nvidia RTX-2080 Ti GPU as well as an Intel Xeon Silver 4210 CPU and the different networks took on the order of two to five days to train. 

\color{black}
\subsection{Metrics}
\label{sec:metrics}
The ensemble of neural networks $\hat{y} = f(x, \theta_j)$ constitute a point cloud approximation of the predictive distribution $p(y \, | \, x, D)$, 
which is 
the full Bayesian solution to the mapping problems. From $p(y \, | \, x, D)$ useful summaries can be computed.
For example, 
we can take the mean and standard deviation of $p(y \, | \, x, D)$
as an estimate of the prediction and the uncertainty in the prediction, respectively. Other commonly used summaries are the mode, median, and credible intervals. 

The main advantage of the Bayesian approach to machine learning models, and neural networks in particular, is that it furnishes an estimate of the uncertainty in the output of a model. 
It is therefore important to assess the reliability of the prediction functions by examining the uncertainty estimates, which come with the predictions. Since we are following a Bayesian approach, it would be natural to use Bayesian measures to assess the results of the models. However, we have chosen to follow common practice in machine learning and assess the reliability of our results using frequentist measures. We do so for two reasons. Firstly, any procedure can be treated as a black box whose performance can be assessed using frequentist methods irrespective of what lies within the box. Secondly, frequentist questions, such as how often a model gets something right, are often easy to answer so long as one makes clear what reference ensemble is being used to make the assessments. Frequentist measures are particularly useful when a calibration of a Bayesian procedure may be necessary\,\cite{Little_2011,bayarri2004} as is the case when aspects of the procedure, such as the choice of prior, are  largely the result of pragmatic considerations rather than formal construction from first principles. 

Our choice of reliability measures is, therefore, informed by the following considerations. We want metrics that are easy to compute, simple to understand, and commonly used to assess the quality of fitted machine learning models.
\color{black}
The quality of the approximations has been
assessed using two main metrics:
\begin{itemize}
    \item the ratio of the absolute difference to the target $y$,  $\textrm{E}[|\hat{y} - y)|] \, / \, y$  and 
    \item the relative frequency with which the 3-standard deviation interval about the mean of $p(y \, | \, x, D)$ brackets the true predictions, $y$. 
\end{itemize}
The second metric provides a
frequentist
calibration of the Bayesian 3-standard deviation 
credible
\color{black}
intervals. We are asking for the confidence level of these intervals, which is a purely frequentist measure, if the credible intervals were to be interpreted as approximate confidence intervals. As noted above, the use of such a metric deviates from pristine Bayesian methodology; nevertheless, it accords with a commonly used approach for assessing statistical results whatever their provenance\,\cite{Little_2011,bayarri2004}. 

The quality of the classifier was assessed using
 the standard machine learning measures precision, recall, and $F_1$\,\cite{10.1007/978-3-540-31865-1_25}. 
 In the current context, the
 precision is given by
\begin{align}
    P & \equiv P(V \, | \, +) \nonumber\\ & =
    \frac{P(+ \, | \, V) \, N(V) }{P(+ \, | \, V) \, N(V) + P(+\,|\, \overline{V}) \, N(\overline{V})},
\end{align}
where $R = P(+ \, | \, V)$ is
the recall and
$N(V)$ and $N(\overline{V})$ are the numbers of valid and invalid pMSSM points, respectively.
A result is $+$ if  a pMSSM point is classified as valid and $-$ otherwise. 
Note that the recall can be
computed from
\begin{align}
    R & \equiv \frac{P(+ \, | \, V) \, N(V) }{P(+ \, | \, V) \, N(V) + P(-\,|\, V) \, N(V)},
\end{align}
where the numerator is the
number of true positives and
the denominator is the sum
of true positives and false
negatives. The quantity $F_1$,
the harmonic mean 
\begin{align}
    F_1 & = \frac{P R}{(P + R)/2},
\end{align}
provides a single number that is a compromise between precision and recall.
Recall is
the fraction of valid points that are correctly identified and
is an intrinsic characteristic of a classifier, whereas precision depends on the ratio $N(V) / N(\overline{V})$, which is a property of the data set to which the classifier is applied. In particle physics, the quantity obtained by setting $N(V) = N(\overline{V})$ in the precision, that is, by using a balanced data set, is typically referred to as the discriminant.

In practice, the precision and recall are
approximated using the
 number of true positive predictions, $TP$, the the number of false positive predictions, $FP$, and the number of false negative predictions, $FN$, as follows,
\begin{align}
    P & =\frac{TP}{TP+FP}, \\
    R & =\frac{TP}{TP + FN}.
\end{align}
For a given input feature vector $x$, here a pMSSM parameter point, a machine learning model would typically provide a single estimate of the quantity it models, while a BNN model provides a point cloud approximation to the full predictive distribution, Eq.\.(\ref{eq:predictive}).  We can assess the impact of using a posterior distribution rather than the output from a single network by noting how the metrics change when applied to the mean, the mean minus 3 standard deviations, and the mean plus 3 standard deviations.


\section{Results}



\begin{figure}
    \centering
    \includegraphics[keepaspectratio, height=0.30\textwidth]{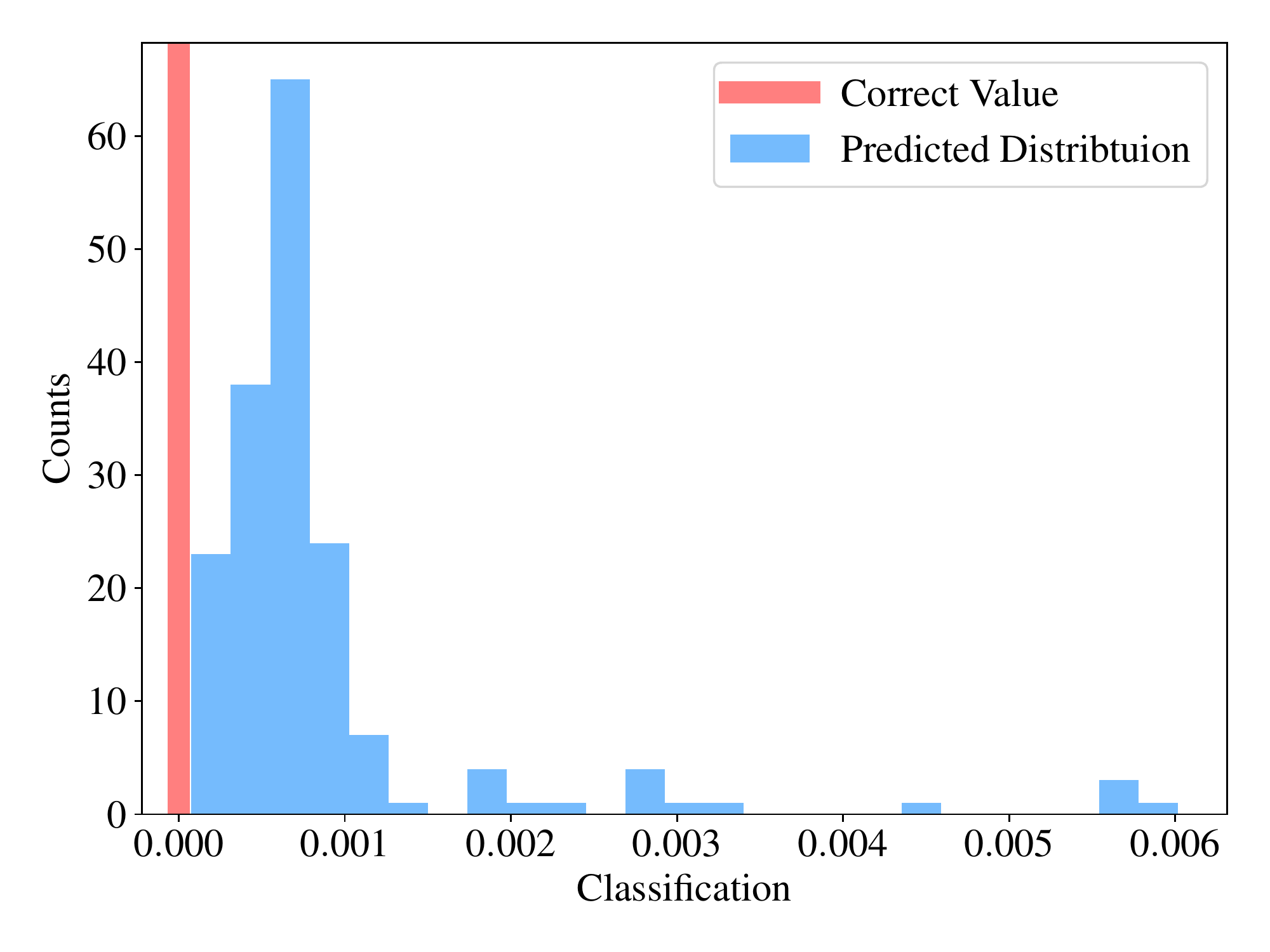}
    \includegraphics[keepaspectratio, height=0.30
    \textwidth]{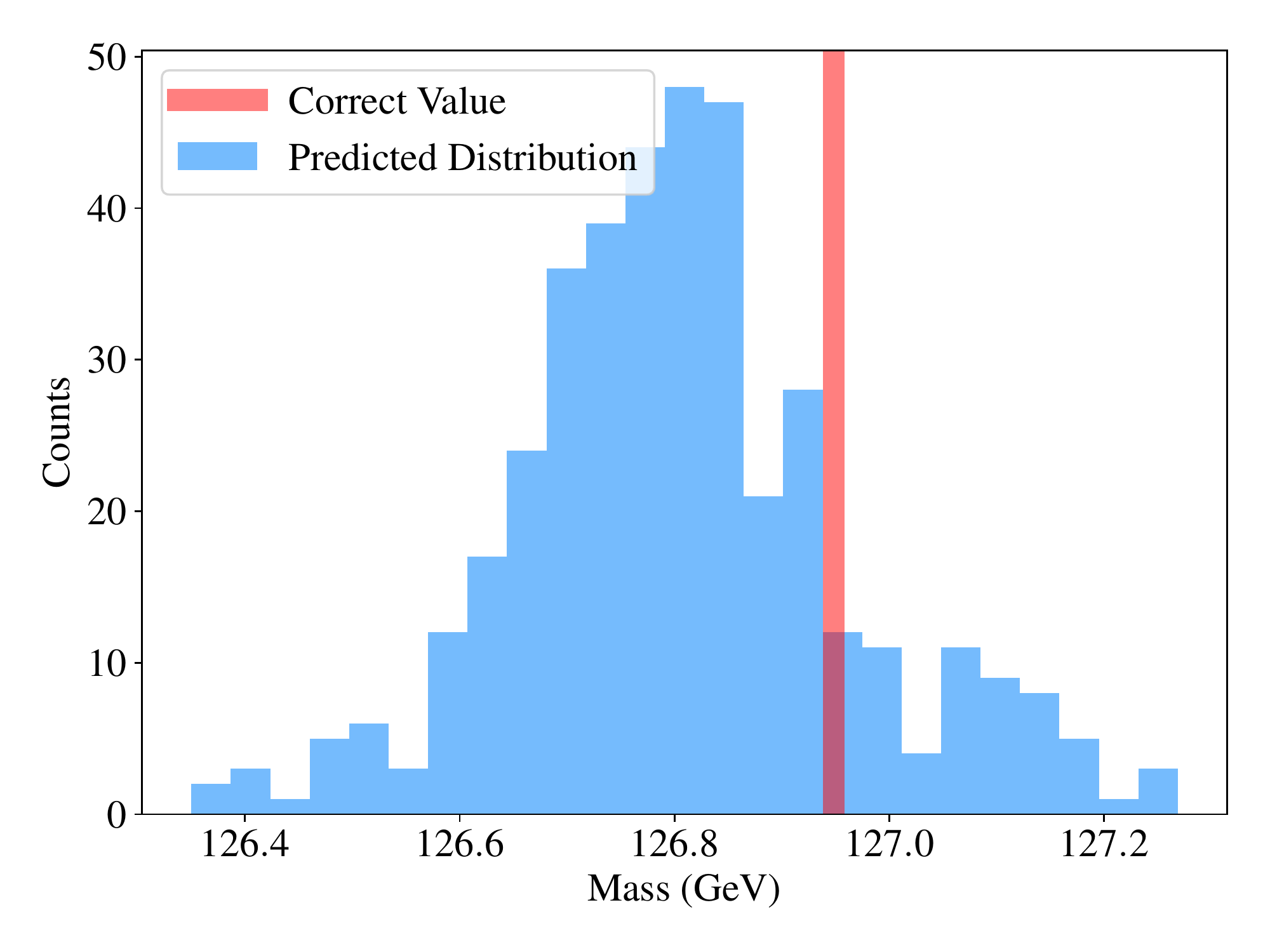}
    \includegraphics[keepaspectratio, height=0.30\textwidth]{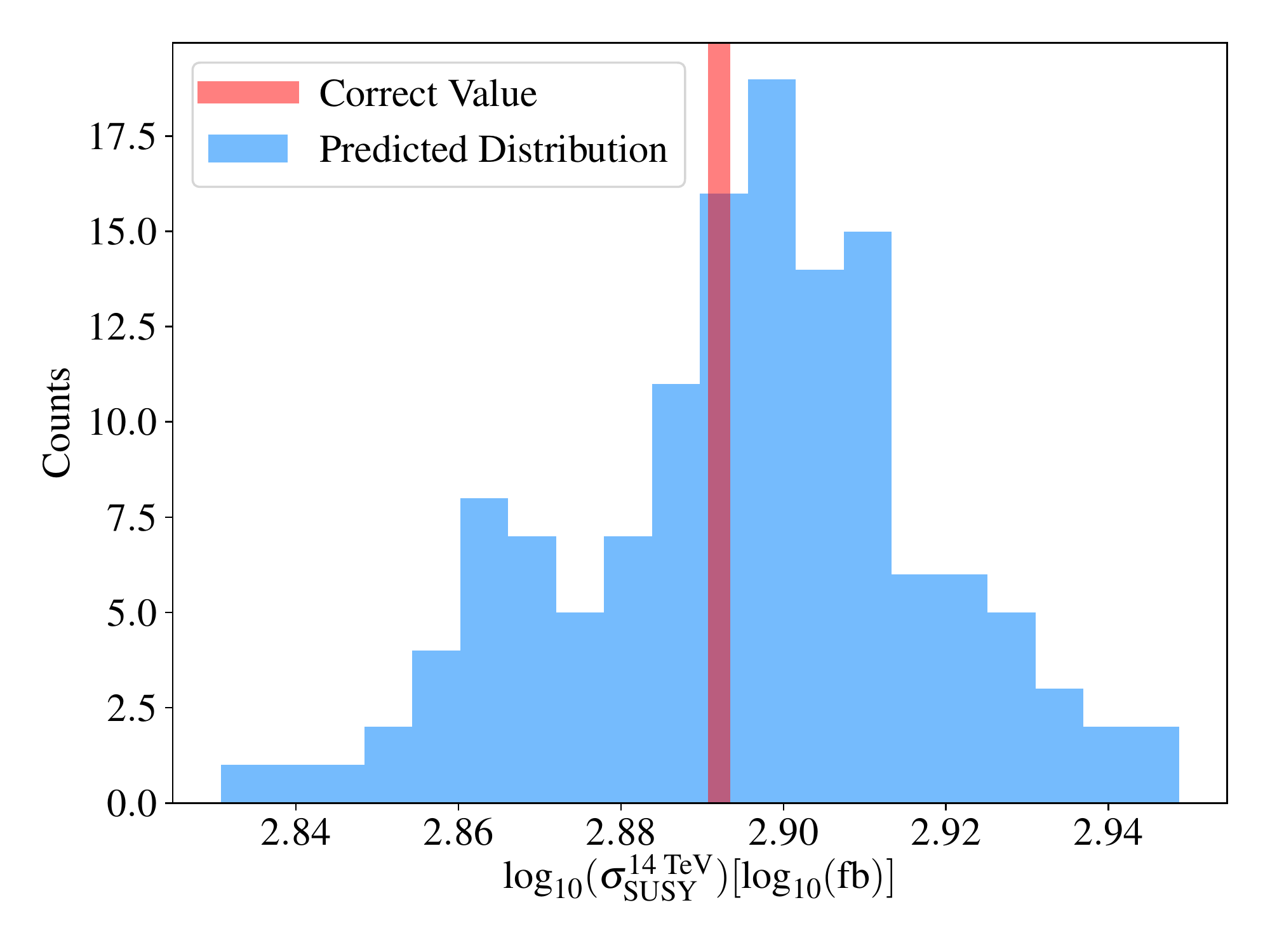}
             \caption{Predictive distributions $p(y | x, \theta)$. (top) classifier; (middle) mass of lightest neural Higgs boson, and (bottom) log cross section.}
    \label{fig:vpar_bayesian_1}
\end{figure}

For each pMSSM prediction function, and a given pMSSM parameter point $x$, we use the mean of the predictive distribution, $p(y \, | x, D)$, as an estimate of the corresponding prediction from either \texttt{SOFTSUSY} or \texttt{Prospino2}, while the standard deviation of the predictive distribution is taken as an estimate of the uncertainty in the prediction.
Figure\,\ref{fig:vpar_bayesian_1} show examples of the predictive distributions for the validity, Higgs boson mass, and log cross section predictions. As stated above, it is these distributions that 
constitute the full Bayesian solutions rather than the summaries, but the summaries are nevertheless useful. Below, we use them to assess the reliability of these distributions. 
\color{black}

\subsection{The viability classifier (VPAR)}

The three performance metrics for the \emph{VPAR} classifier are given in Table \ref{tab:vparTrain}. A pMSSM point is classified as valid if the point estimate, the average over the ensemble of networks, exceeds a cutoff of 0.5. 
 Alternatively, one can apply the cutoff to the average plus 3 standard deviations.
This causes many points with high uncertainty to be classified as valid and results in a much higher recall score and slightly lower $F_1$ and precision scores. We expect the average over the ensemble of networks to be most useful for attaining the highest classification accuracy, while the average plus 3 standard deviation would be useful in selecting valid points that could serve as input to a program such as \texttt{SOFTSUSY}. 
The program would catch false positives but not false negatives. 

\begin{table}[ht]
\begin{center}

\begin{tabular}{ | c | c | c | c |} 
\hline
Configuration & Recall & Precision & F1 \\ 
\hline
Mean & 0.955 & \textbf{0.953} & \textbf{0.954} \\ 
\hline
Mean + 3SD & \textbf{0.982} & 0.915 & 0.947 \\ 
\hline
\end{tabular}
\end{center}

\caption{\textbf{VPAR}: performance metrics.}
\label{tab:vparTrain}
\end{table}

\subsection{The cross section regression function (XSEC)}
After training on the \textbf{VPAR} data set, networks were trained on the \textbf{XSEC} data set.  For a given pMSSM parameter point, we again take  the average, $\hat{y}$, of the distribution of network outputs, as an estimate of the quantity being modeled, here the predicted cross section in femtobarns. We use the associated standard deviation, $\sigma$, to exclude pMSSM parameter points 
for which $\log(\hat{y} - 3 \sigma) > 3$, as pMSSM points with cross section of this size or larger have been excluded at the LHC\,\cite{ATLASpMSSM,Khachatryan:2016nvf}. This cut removed 5.5$\%$ of the generated pMSSM points. On average, with this cut, the cross section is estimated with a percentage uncertainty of 3.34$\%$ and the true value fell within 3 standard deviations of the estimated cross section $99\%$ of the time. 

The standard deviation can be used to flag pMSSM points for which the network based prediction is highly uncertain. If we exclude the 45 pMSSM points with cross sections that differ by an order of magnitude or more between the bounds $\hat{y} - 3 \sigma$ and $\hat{y} + 3 \sigma$ the relative uncertainty in the predictions falls to 3.04$\%$. 

\begin{figure}
    \centering
    \includegraphics[keepaspectratio, height=10cm]{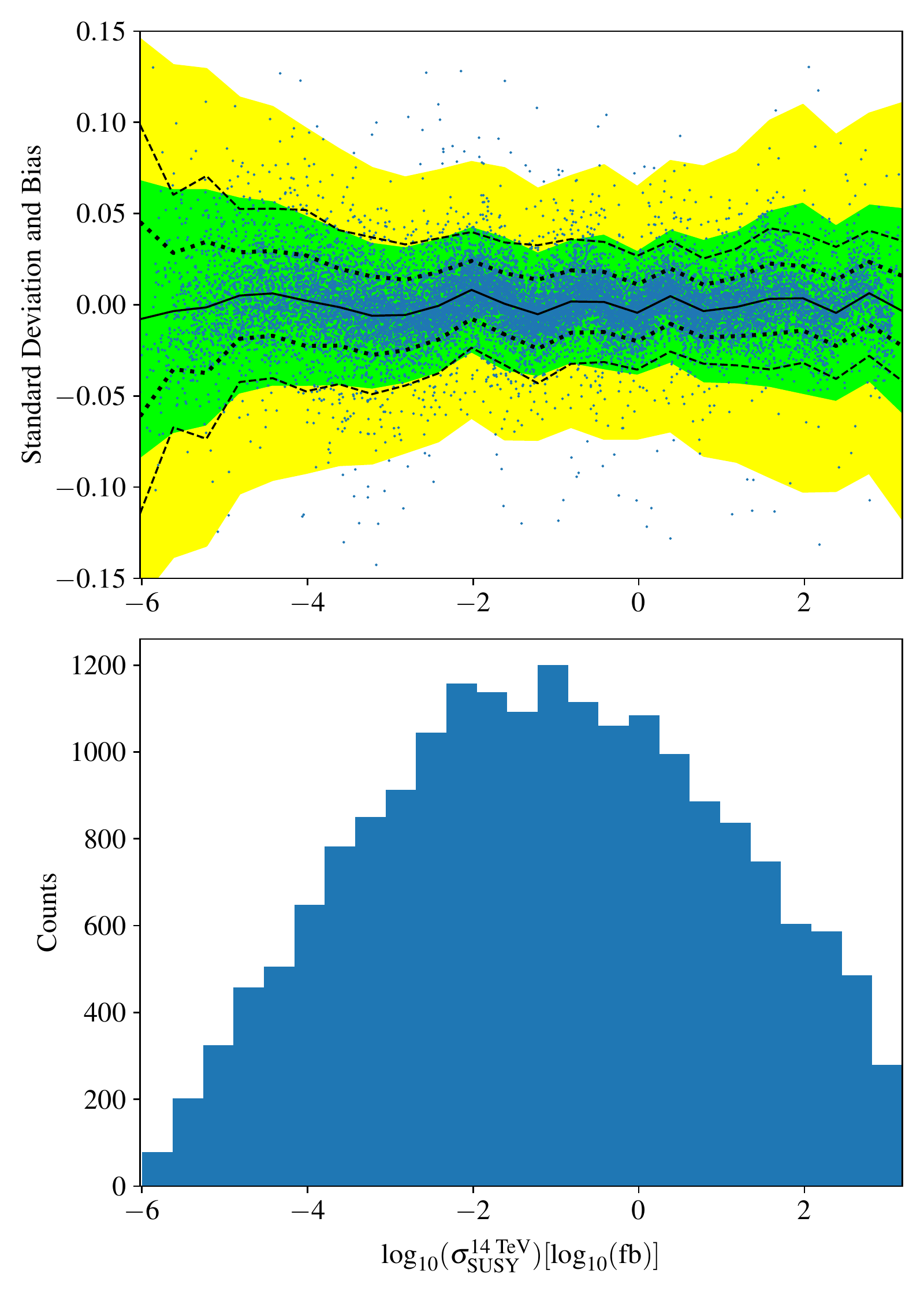}
     \caption{(top) The solid and dotted lines are computed from the errors of the BNN, that is, from the differences between the BNN estimates and the associated true values. The black line is the bias of the BNN as a function of the $\log$ of the cross section, while the dotted lines represent the 1 and 2 standard deviation intervals. The green and yellow bands are root mean square intervals computed from the standard deviations furnished by the BNN.  (bottom) The distribution of the log cross section.}
    \label{fig:XSEC_Brazil}
\end{figure}

Figure\,\ref{fig:XSEC_Brazil} shows the two measures of uncertainty in the BNN predictions: one computed directly from the known errors of the BNN predictions and the other from the estimated standard deviations of the predictive distribution. The bias in the BNN predictions is negligible. However, the point cloud of network outputs overestimates the uncertainty in the BNN predictions. It is also clear that the BNN behaves as expected in that the uncertainty is greater where there are fewer data for training.

When predicting cross sections for many pMSSM points simultaneously on a GPU, a single prediction per network, in the ensemble of networks, took on average 49 nanoseconds for the prediction alone and 94 nanoseconds when the computational overhead is included. The BNN for the above results used an ensemble of 135 networks, which implies a computation rate of 12.7 microseconds per prediction including the overhead. This is approximately 16.5 million times faster than running \texttt{Prospino2}, which took about 3.5 minutes per NLO prediction.

Sampling via any Markov chain Monte Carlo method, including Hamiltonian Monte Carlo, produces a sequence of correlated predictions. While one may estimate mean quantities using such predictions, estimating uncertainties from these points requires more care because of the correlation. It is usually simpler to down sample to a set of approximately \emph{iid} points by using every $n$ points along a chain, where, ideally, the choice of the integer $n$ is somewhat longer than the correlation length along the chain. 
For example, the ensemble size of 135 
for the cross section prediction
function is a compromise between the desire to have an ensemble large enough to produce distributions such as those in Fig.\,\ref{fig:vpar_bayesian_1} and small
enough to permit very fast predictions.

\color{black}

\begin{figure}

    \centering
    \includegraphics[keepaspectratio,height=4
    cm]{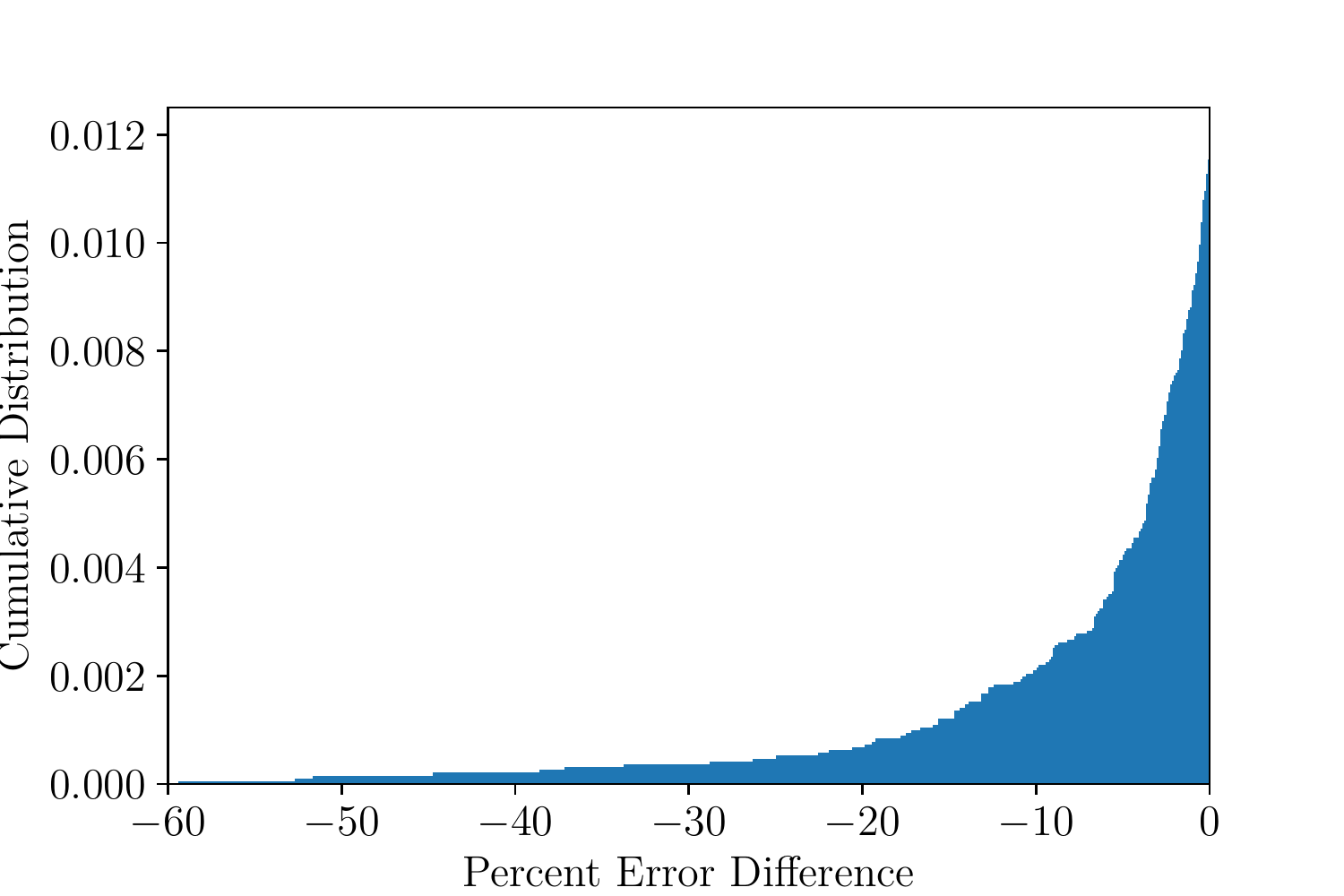}
    \includegraphics[keepaspectratio,height=4cm]{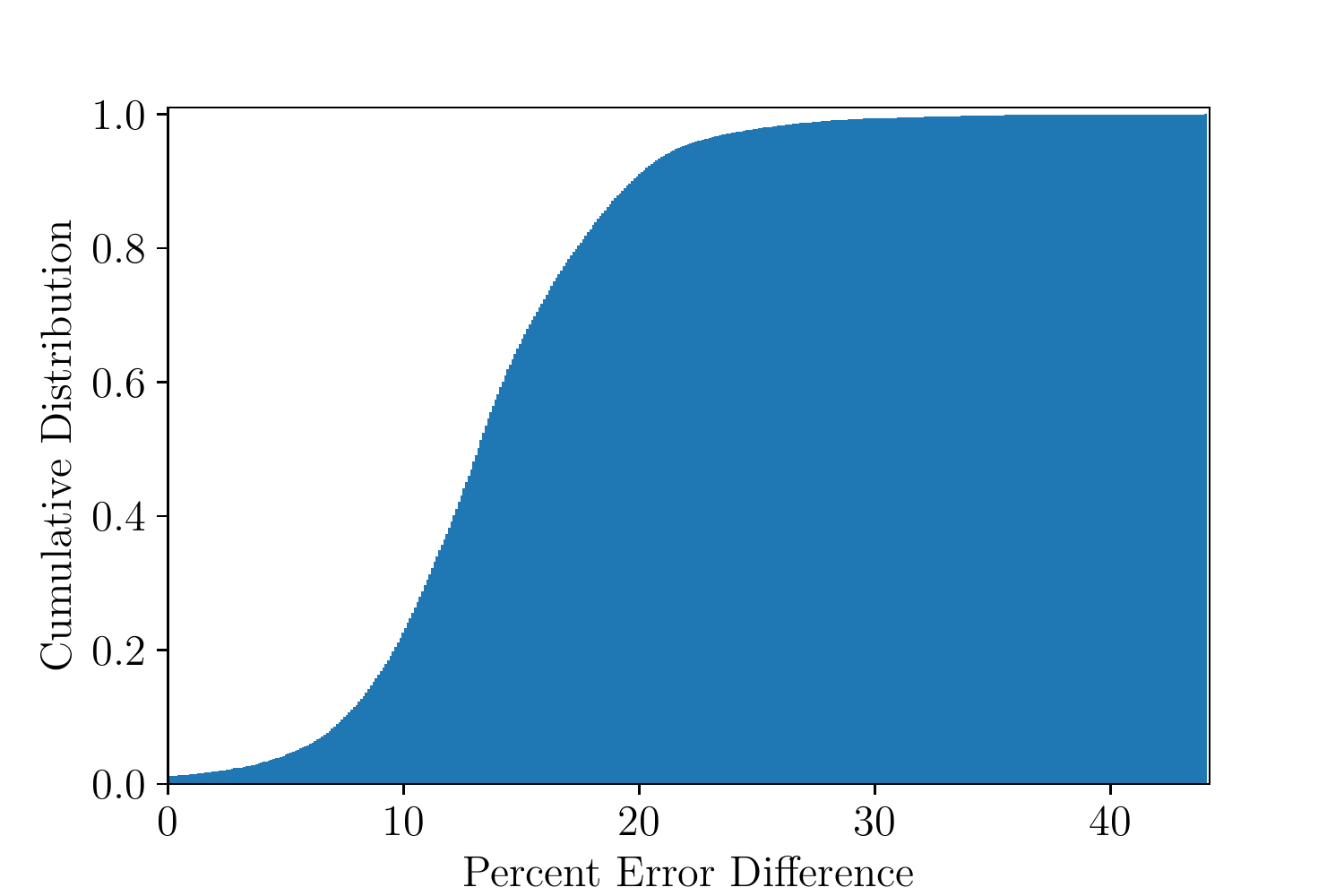}
    \caption{The cdf for the difference in relative uncertainty between (top) the prediction at leading order and that of the BNN and (bottom) between the BNN and the leading order. The vast majority of the BNN-based predictions are more precise, in the sense of being closer to the corresponding NLO predictions, than the predictions at leading order. }
    \label{fig:CDF}
\end{figure}

When cross sections are required for a large number of pMSSM points, the leading order prediction is frequently used as it is faster to compute. However, on average, it is approximately $19\%$ less precise than the prediction at next-to-leading order (NLO) accuracy, while, as noted above, on average the BNN matches the NLO prediction within $3.0\%$ when the few poorly estimated cross section predictions are excluded. Note, however, that even after excluding these outliers, we find that there are still a number of points for which the BNN does worse than the leading order prediction. But, as can be seen in Figure \ref{fig:CDF}, this happens for a small percentage of the pMSSM points we considered.

\subsection{The Higgs boson mass function (OHIGGS)}

The output of the \textbf{OHIGGS} data set was analyzed using both regression and classification approaches with the same ensemble of networks. For both analyses the data set was split into two subsets. One consisted of all pMSSM points where either the true Higgs boson mass was within 2\,GeV of 125\,GeV\,\cite{201230}, or the point's 3 standard deviation interval overlapped with this range. The other subset consisted of all remaining points. 
Within the first subset, the average percent error was $0.10\%$ and $87.4\%$ of the time the true value was within 3 standard deviations of the predicted value. In the second subset, the percent error was $0.14\%$ and $86.7\%$ of the time the credible interval contained the true value. If the predictive distributions were Gaussian, these credible intervals, if interpreted as approximate confidence intervals, undercover by about $13\%$, which shows they are not as well calibrated as the ones for the cross section data set.

In the classifier approach, the labels were positive if the true Higgs boson mass was within 2\,GeV of 125\,GeV, and negative if it was not. We claim a prediction to be positive, that is, good, if any part of its credible interval overlapped the desired range. Using this classification criterion, the precision of the network was 0.926, its recall was 0.997, and its $F_1$ score was 0.960. We therefore conclude that using the BNN to identify pMSSM parameter points with a low-mass neutral Higgs boson consistent with the measured Higgs boson mass will remove very few pMSSM points that yield Higgs boson masses consistent with observation. Using this approach in conjunction with the regression will also allow very accurate labelling of the selected masses as its error on the selected pMSSM points is low. 

\begin{figure}
    \centering
    \includegraphics[keepaspectratio, height = 10cm]{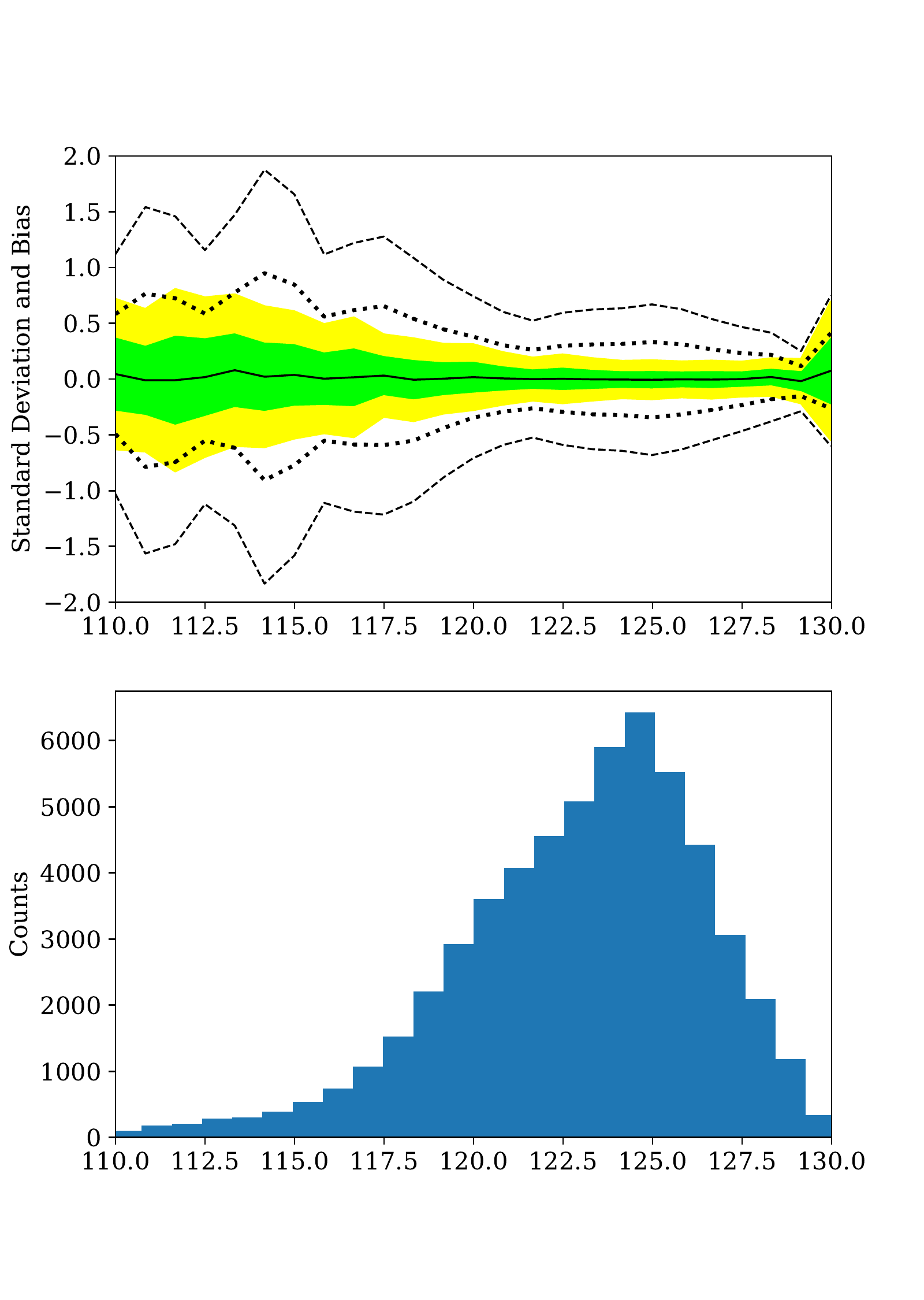}
    \caption{(top) The solid and dotted lines are computed from the errors of the BNN. The black line is the bias of the BNN as a function of the Higgs boson mass. The dotted lines represent the 1 and 2 standard deviation intervals. The green and yellow bands are root mean square intervals computed from the standard deviations furnished by the BNN.  (bottom) The distribution of the Higgs boson mass.}
    \label{fig:OHIGGS_Brazil}
\end{figure}

We see in Figure \ref{fig:OHIGGS_Brazil} that the estimated uncertainty from the ensemble of networks does not match the uncertainty computed from the actual errors. Moreover, in contrast to the results for the cross section, the Higgs boson mass BNN  underestimates the uncertainties, though, as expected, the uncertainties are larger where there are fewer data. 

\subsection{Discussion}
The utility of Bayesian neural networks is that they directly approximate the predictive distribution $p(y \, | \, x, D)$, that is, they provide a probability density over the space of the quantity being modeled. Moreover, their implementation on GPUs yields a significant increase in prediction speed. With the GPUs used, we achieve a computation speed about 50,000 times faster than \texttt{SOFTSUSY} and 16.5 million times faster than \texttt{Prospino2} running on a single CPU.

In principle, the predictive distribution encodes the uncertainty in a given BNN prediction. 
There are two kinds of uncertainty that should be accounted for. The first is the uncertainty arising from the fact that a finite amount of data are used to train, that is, fit the models. The second is the uncertainty due to the fact that we do not know which model should be fitted. However, to the degree that the neural network models used in this paper are sufficiently flexible, the uncertainties reported in this paper automatically include both. 

However, as is true of all Bayesian inference, the results depend on the likelihood function as well as on the prior. In this work, we have chosen the form of the prior for computational simplicity, with parameters constrained by hyper-priors for added flexibility. However, even granting the form of the hierarchical prior, it is still necessary to choose the values of the hyper-parameters. We have made no attempt, so far, to optimize those choices, which may explain both the over and under estimates of the standard deviations associated with the BNN predictions. But the fact that the problem depends upon hyper-parameters as is true of all machine learning models can be turned into a virtue. For example, by weighting the output of each network in the ensemble of networks by the ratio of the hyper-prior with its hyper-parameters viewed as variables to the hyper-prior with which the HMC sampling was done, we may be able to use standard optimization techniques to improve the results by optimizing the choice of hyper-parameters.



\section{Conclusion}
Given a large number of predictions of interest from high-dimensional models such as the pMSSM, we showed that BNNs, implemented on GPUs, can successfully model these predictions with computation speeds from 50,000 to 16.5 million times greater than that of the programs that yielded the predictions. This makes it possible to make rapid, accurate, predictions for model points other than those used to construct the BNNs. In particular, we were able to classify accurately whether a given pMSSM parameter point is valid, as determined by \texttt{SOFTSUSY},  with a maximum $F_1$ score of 0.957 
and a recall of 0.987. 
We were also able to predict the cross sections for the production of supersymmetric particles that matched the predictions at NLO accuracy to about $3\%$ on average. 
Finally, we could classify whether pMSSM parameters would give a Higgs boson mass of $125\pm 2$\,GeV with a recall of 0.998 and an 
$F_1$ score of 0.942.

These results indicate that it will be possible to more easily filter out pMSSM parameter combinations that are either unphysical or that predict values for the Higgs boson mass inconsistent with observation.  
It will also be much easier to study the impact of varying different parameters on the production cross sections of supersymmetric particles. Finally, the predictions studied in this paper were just a small number of the potentially interesting ones. Using the methods described in this paper, which are available in the \texttt{TensorBNN}\,\cite{TensorBNN} package, it should be possible to apply these methods to make rapid predictions of other SUSY observables given a large sample of these predictions from programs such as \texttt{SOFTSUSY} and \texttt{Prospino2}. Moreover, as noted in our discussion, there is much room for improvement.


\section{Acknowledgements} This work was supported in part by the Davidson Research Initiative and the U.S. Department of Energy Award No. DE-SC0010102.

\section*{}



\bibliography{main.bbl}

\end{document}